\documentclass[sigplan,nonacm]{acmart}
\AtBeginDocument{%
  }

\setcopyright{acmlicensed}
\copyrightyear{2018}
\acmYear{2018}
\acmDOI{XXXXXXX.XXXXXXX}
\acmConference[Conference acronym 'XX]{Make sure to enter the correct
  conference title from your rights confirmation email}{June 03--05,
  2018}{Woodstock, NY}
\acmISBN{978-1-4503-XXXX-X/2018/06}

\usepackage{graphicx}
\usepackage{subcaption}
\usepackage{float}
\usepackage{listings}
\usepackage{amsmath}
\usepackage{xcolor}
\usepackage{tabularx}
\usepackage{tabularray}
\usepackage{array}
\usepackage{makecell}
\definecolor{codegreen}{rgb}{0,0.6,0}
\definecolor{codegray}{rgb}{0.5,0.5,0.5}
\definecolor{codepurple}{rgb}{0.58,0,0.82}
\definecolor{backcolour}{rgb}{1,1,1}

\lstdefinestyle{mystyle}{
    backgroundcolor=\color{backcolour},   
    commentstyle=\color{codegreen},
    keywordstyle=\color{magenta},
    numberstyle=\tiny\color{codegray},
    stringstyle=\color{codepurple},
    basicstyle=\ttfamily\footnotesize,
    breakatwhitespace=false,         
    breaklines=true,                 
    captionpos=b,                    
    keepspaces=true,                 
    numbers=left,                    
    numbersep=5pt,                  
    showspaces=false,                
    showstringspaces=false,
    showtabs=false,
    frame=single,
    rulecolor=\color{black},
    tabsize=2
}

\usepackage{pifont}

\newcommand{\system}{Kairos}

\begin{document}


\title{\system{}: Low-latency Multi-Agent Serving with Shared LLMs and Excessive Loads in the Public Cloud}





\author{Jinyuan Chen$^{*}$, Jiuchen Shi$^{*}$, Quan Chen$^\dagger$, Minyi Guo}

\affiliation{
  \institution{Shanghai Jiao Tong University}
  \city{}
  \country{}
}
\thanks{$^*$Co-first authors. \quad $^\dagger$Corresponding author.}
\renewcommand{\shortauthors}{J. Chen et al.}


\begin{abstract}
Multi-agent applications utilize the advanced capabilities of large language models (LLMs) for intricate task completion through agent collaboration in a workflow.
Under this situation, requests from different agents usually access the same shared LLM to perform different kinds of tasks, forcing the shared LLM to suffer excessive loads.
However, existing works have low serving performance for these multi-agent applications, mainly due to the ignorance of inter-agent latency and resource differences for request scheduling.

We therefore propose \textbf{\system{}}, a multi-agent orchestration system that optimizes end-to-end latency for multi-agent applications.
\system{} consists of a \textit{workflow orchestrator}, a \textit{workflow-aware priority scheduler}, and a \textit{memory-aware dispatcher}.
The orchestrator collects agent-specific information for online workflow analysis.
The scheduler decides the serving priority of the requests based on their latency characteristics to reduce the overall queuing.
The dispatcher dispatches the requests to different LLM instances based on their memory demands to avoid GPU overloading.
Experimental results show that \system{} reduces end-to-end latency by 17.8\% to 28.4\% compared to state-of-the-art works.
\end{abstract}

\maketitle

\section{Introduction}
Multi-agent applications leverage the advanced capabilities of large language models (LLMs) in language understanding and generation to achieve enhanced quality in complex task execution through role specialization and collaboration~\cite{hong2024metagpt,zhou2023agentsopensourceframeworkautonomous}. These applications decompose complex tasks into structured, multi-stage sub-tasks in a workflow~\cite{qian2024chatdev,wu2023autogenenablingnextgenllm,langflow,gpt_newspaper}, which are collaboratively completed by LLM-driven agents with different responsibility boundaries that are differentiated through specialized system prompts~\cite{hong2024metagpt,qian2024chatdev,gpt_researcher}. 
For instance, a Question Answer (QA) application includes the ``Router'', ``Humanities'', and ``Math'' agents to cooperatively answer various types of questions~\cite{gao2024agentscope}.

Due to LLMs' general-purpose and multi-task capabilities, multiple agents usually utilize the same LLM~\cite{10.5555/3691938.3691988,10.1145/3676641.3716278,hong2024metagpt,wu2023autogenenablingnextgenllm}.
However, our observations reveal that different agents inherently exhibit obvious differences in LLM execution characteristics, leading to varying execution latency and GPU memory demands. 
For example, in the QA, the ``Math'' agent generates answers with significantly longer LLM outputs than the ``Router'' agent, which only performs a quick routing decision, with the latency variances reaching up to 25.1X. Moreover, the agents' specific positions within the multi-agent workflow lead to varying remaining execution latency.

\begin{figure}
  \centering
  \includegraphics[width=0.78\linewidth]{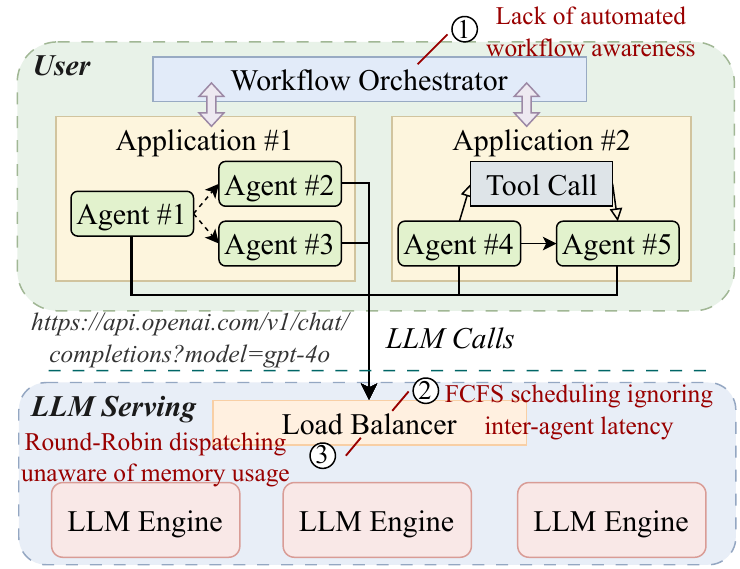}
    \caption{Multi-Agent application serving architecture.}
  \label{fig:AgentInfrastructure}
  \vspace{-6mm}
\end{figure}

Figure~\ref{fig:AgentInfrastructure} shows the Multi-Agent application serving architecture.
We can see agents within/across different applications will utilize the same LLM deployed by the LLM provider.
The LLM provider utilizes a workflow orchestrator to manage task coordination among agents. Moreover, it utilizes a load balancer to manage the serving of requests generated by different agents.
With the highly shared LLM paradigm, the workloads for an LLM will be excessive and more dynamic~\cite{BurstGPT,QLM} in public cloud serving.
When the online serving loads dynamically change to be excessive, lots of queries from different agents can be queued at the shared LLM before elastic resource scaling is completed~\cite{QLM,alibaba_cloud,gke}. 

Under this situation, it is desirable to reduce the end-to-end latency for the Multi-Agent application.
Therefore, at the load balancer, efficient request priority scheduling based on remaining execution latency is required to reduce the overall queuing time. Also, queries need to be appropriately dispatched to LLM instances based on their GPU memory demands.   
Moreover, such scheduling and dispatching require awareness of agent-specific execution characteristics and application workflow context at the workflow orchestrator.

Existing works~\cite{10.5555/3691938.3691988,10.1145/3676641.3716278,kwon2023efficient,298679,298687} fall short of achieving the above goals for multi-agent application serving. Under conditions of high load and significant request queuing, they exhibit high end-to-end latency due to their neglect of differences among agents in execution latency and memory demand.
From our evaluations, the latency per token can be as high as 2.7s.
The specific problems of existing works primarily stem from the following three aspects.

As for the first problem (Figure~\ref{fig:AgentInfrastructure} \textcircled{1}), they lack automated workflow analysis mechanisms. 
For efficient multi-agent serving, it is crucial to obtain the workflow structure to determine each agent's position and its remaining execution latency. 
However, existing workflow analysis approaches are inherently static, requiring developers to either explicitly define the entire workflow structure or expose data dependencies via APIs. Although having consumed lots of manual efforts, this static reliance cannot capture the dynamic and uncertain call relationships of autonomously planning multi-agent applications, thus failing to parse these workflows.

As for the second problem (Figure~\ref{fig:AgentInfrastructure} \textcircled{2}), current works ignore inter-agent remaining latency differences for request scheduling. 
Without the awareness of requests' end-to-end remaining execution latency, they cannot identify which requests are closer to completion. 
Therefore, when requests from different agents wait in the queue under excessive loads, current works simply utilize the First-Come-First-Serve (FCFS) for scheduling~\cite{kwon2023efficient,10.5555/3691938.3691988}.
This results in requests with short remaining latency being queued behind those with longer remaining latency, causing request-level head-of-line blocking and significantly increasing overall queuing.

As for the third problem (Figure~\ref{fig:AgentInfrastructure} \textcircled{3}), 
current works are unaware of inter-agent GPU memory usage differences for request dispatching. 
Popular LLM engines~\cite{kwon2023efficient,zheng2024sglangefficientexecutionstructured} typically employ continuous batching and dynamic memory allocation to improve throughput.
Under these strategies, when failing to recognize the differences in memory demands among agents,  
suboptimal request batching on LLM instances can be formed, which can cause some GPUs to be overloaded while others are idle. The requests scheduled to overloaded GPUs will be frequently preempted and thus need to be recomputed. Preempted requests waste already invested resources and interfere with the normal execution of other requests.

To address the above issues, we develop a flexible Multi-Agent orchestration system, \textbf{\system{}}.
\system{} consists of a \textit{workflow orchestrator}, a \textit{workflow-aware priority scheduler}, and a \textit{memory-aware time-slot dispatcher}.
The orchestrator collects historical execution information of LLM requests online, enabling automatic workflow analysis and modeling of execution characteristics for requests of different agents. 
The scheduler perceives the remaining execution latency among requests of different agents, allowing it to prioritize requests that will complete sooner, to reduce overall queuing latency. 
The dispatcher leverages the differences in memory demands among requests of different agents, enabling it to dispatch requests to LLM instances with the most suitable available GPU memory, to avoid request preemption. 
The main contributions of this paper are as follows:

\begin{itemize}
    \item \textbf{Investigating the inefficiencies of the current Multi-Agent request scheduling.} The analysis identifies the significance of collecting diverse inter-agent execution latency and GPU memory demands and including them in the query scheduling and dispatching. 
    \item \textbf{The design of the workflow orchestrator.} The orchestrator automatically collects critical agent-specific information to enable efficient request scheduling.
    \item \textbf{The design of the workflow-aware scheduler.} The scheduler perceives the remaining latency of requests across agents, thereby optimizing priority decisions and significantly reducing overall queuing latency.
    \item \textbf{The design of the memory-aware dispatcher.} The dispatcher perceives the memory demands among requests of different agents and dispatches them to the LLM instance with the most suitable available memory, thereby enabling efficient request batching.
\end{itemize}

We implement \system{} on top of vLLM~\cite{kwon2023efficient}, and employ Kafka message queue~\cite{kafka} for inter-agent communication to support distributed deployment and request scheduling. We select three representative multi-agent benchmarks~\cite{wu2023autogenenablingnextgenllm,hong2024metagpt,gao2024agentscope} to evaluate \system{} on 4 NVIDIA A40 GPUs. 
Experimental results show that \system{} reduces the end-to-end latency from 17.8\% to 28.4\% compared to state-of-the-art works.
\section{Motivation}
In this section, we first characterize the inter-agent differences in LLM inference and then expose the limitations of existing works. At last, we conclude the design requirements for efficient multi-agent deployment.

\subsection{Inter-Agent Differences in LLM Inference}

\subsubsection{Representative Multi-Agent Applications}
To gain a comprehensive understanding of the characteristics of real-world multi-agent applications, we investigated prominent open-source multi-agent projects available on GitHub using the keyword "LLM Multi-Agent", selecting 30 best-matched projects with more than 1,000 stars. This investigation revealed three major types of multi-agent workflows: Dynamic branching, Sequential execution, and Dynamic feedback, as presented in Table~\ref{tab:workflow_survey}. 

\begin{table}[htbp]
    \vspace{-3mm}
    \centering
    \caption{Statistics of representative multi-agent workflows.}
    \label{tab:workflow_survey}
    \begin{tabular}{l|c|c}
        \hline
        \textbf{Workflow Type} & \textbf{Count} & \textbf{Proportion} \\
        \hline
        Dynamic branching & 19 & 63.3\% \\
        Sequential execution & 23 & 76.6\% \\
        Dynamic feedback & 16 & 53.3\% \\
        \hline
    \end{tabular}
\end{table}

Dynamic branching~\cite{gao2024agentscope,agno,langroid} is characterized by adaptive execution paths based on runtime conditions. Sequential execution~\cite{wu2023autogenenablingnextgenllm,lazyllm,chen2024mindsearch} involves a series of pre-defined, ordered steps. Dynamic feedback~\cite{hong2024metagpt,qian2024chatdev,autoagents} incorporates iterative refinement loops where agents re-evaluate and adjust their actions based on previous outputs.

\begin{figure}
  \centering
  \subcaptionbox{Question Answer (QA)\label{fig:benchmark-qa}}{%
    \includegraphics[height=3.5cm]{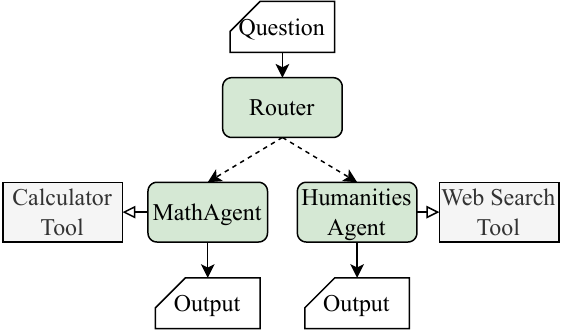}}%
    \hspace{2em}
  \subcaptionbox{Report Generate (RG)}{%
    \includegraphics[height=4cm]{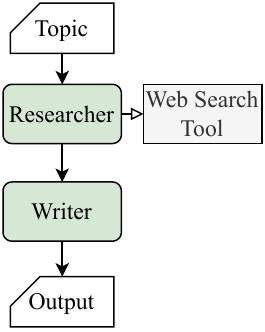}}%
    \hspace{2em}
  \subcaptionbox{Code Generate (CG)}{%
    \includegraphics[height=3.5cm]{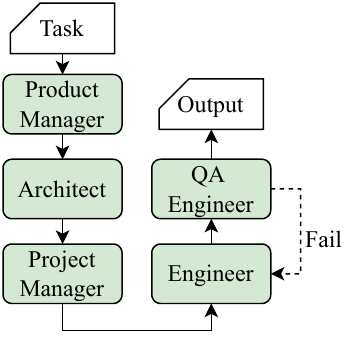}}%
  
  \caption{The three benchmarks represent typical multi-agent applications with different workflow structures.}
  \label{fig:benchmark}
  \vspace{-4mm}
\end{figure}

Based on the prevalence observed in our survey, we selected representative applications for each of these workflow types: Question Answer (QA)~\cite{gao2024agentscope}, Report Generate (RG)~\cite{wu2023autogenenablingnextgenllm}, and Code Generate (CG)~\cite{hong2024metagpt}. Their workflow structures are illustrated in Figure~\ref{fig:benchmark}.

\subsubsection{Benchmarks and Datasets Description}
\label{sec:motivation_description}

We further detail the three applications illustrated in Figure~\ref{fig:benchmark} along with the datasets used for their evaluation.

\textit{Question Answer (QA)} represents a dynamic branching workflow. After a user submits a question, a routing agent determines its category (mathematics or humanities) and assigns the task to the corresponding expert agent for answering. To construct mixed datasets covering both mathematics and humanities questions, we select three mathematics datasets (GSM8K~\cite{gsm8k}, MathQA~\cite{mathqa}, SVAMP~\cite{svamp}) and three humanities datasets (the history subset of MMLU~\cite{mmlu}, WorldHistoryQA~\cite{world_history_1500_qa}, SocialIQA~\cite{socialiqa}). These datasets are paired in equal proportions to:
GSM8K + MMLU (G+M), MathQA + WorldHistoryQA (M+W), SVAMP + SocialIQA (S+S).

\textit{Report Generate (RG)} represents a sequential execution workflow. Given a research topic from the user, a research agent generates materials, followed by a writer agent that generates the final report. The application is evaluated using three datasets as the inputs of research questions or topics: TruthfulQA (TQ)~\cite{truthfulqa}, News Category Dataset (NCD)~\cite{newscategorydataset}, and Natural Questions (NQ)~\cite{naturalquestion,naturalquestion_open}.

\textit{Code Generate (CG)} represents a dynamic feedback workflow. Upon receiving a code development task from the user, multiple roles, including product manager, architect, project manager, engineer, and QA engineer, collaborate to complete the development task. If code evaluation fails, the task is fed back to the engineer for redevelopment, forming a dynamic iterative feedback loop. The application is evaluated using three code generation benchmark datasets: HumanEval (HE)~\cite{humaneval}, MBPP~\cite{mbpp}, and APPS~\cite{hendrycks2021apps}.

\subsubsection{Analysis of Inter-Agent Differences}
\label{sec:motivation_character}

\begin{figure}
  \centering
  \includegraphics[width=0.9\linewidth]{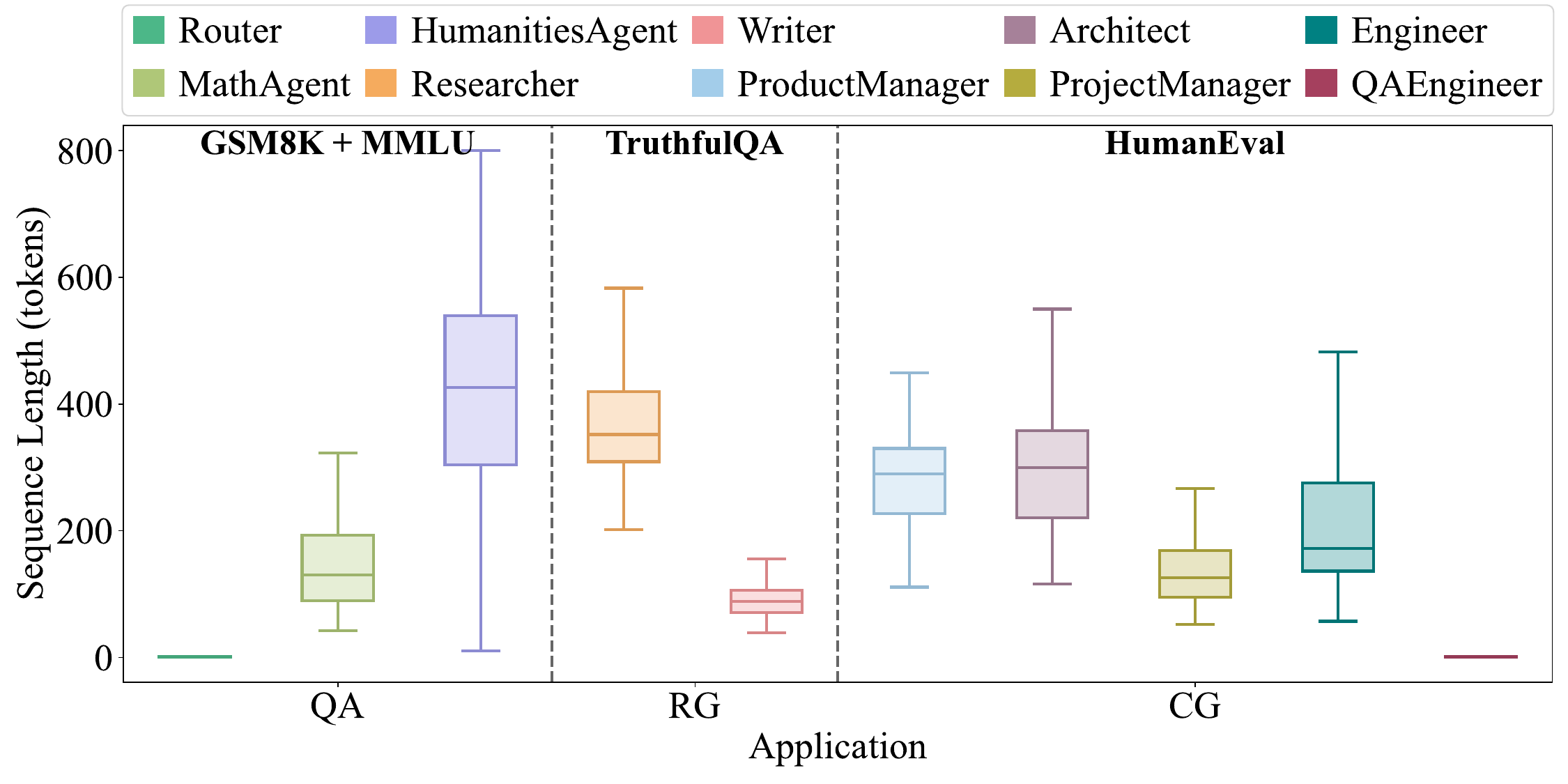}
  \caption{Output length distributions of different agents.}
  \Description{Box plots showing output length distributions for agents in QA, RG, and CG multi-agent applications.}
  \label{fig:combined_box_tokens}
  \vspace{-3mm}
\end{figure}
\begin{figure}
  \centering
  \includegraphics[width=0.9\linewidth]{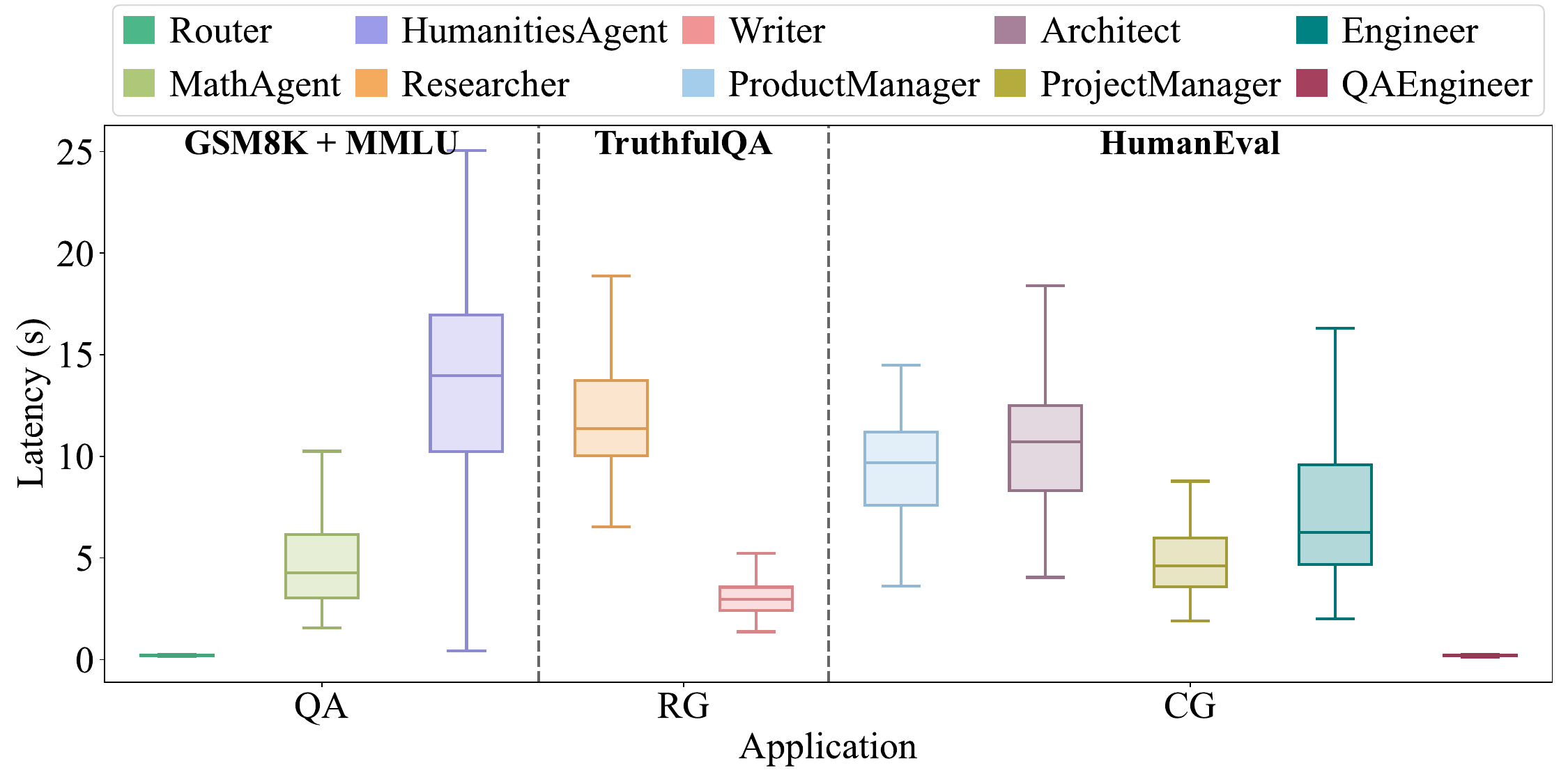}
  \caption{Inference latency distributions of different agents.}
  \Description{Box plots of LLM inference latency distributions for agents in QA, RG, and CG applications.}
  \label{fig:combined_box_latency}
  \vspace{-3mm}
\end{figure}

In multi-agent applications, agents serve different roles, leading to varied LLM inference performance. As shown in Figure~\ref{fig:benchmark-qa}, the QA app includes a Router, Humanities, and Math agent. The Router gives short responses by directing queries to the right agent. The Humanities agent answers open-ended questions with long, structured text, while the Math agent solves problems with brief, formula-based replies. These role differences cause a large variation in output lengths.

To validate the above observation, we analyze the LLM execution in three multi-agent applications: QA with G+M dataset, RG with TQ dataset, and CG with HE dataset. We evaluate them by using the Llama3-8B model on an NVIDIA A40 GPU.
Figure~\ref{fig:combined_box_tokens} shows output length distributions across agents.
We can observe notable differences in output lengths among the ten agents, both within one or across different applications. These results highlight the requirement for public LLM providers to manage diverse execution behaviors from heterogeneous agent requests.

We further analyze the LLM inference latency of these requests. As shown in Figure~\ref{fig:combined_box_latency}, we can observe that the output length directly affects decoding latency, leading to significant latency differences across agents. Since the prefill stage is much faster than decoding, it contributes little to total inference time, with over 96.6\% coming from decoding. These results show that agent functions cause corresponding variations in inference latency.

\begin{figure}
  \centering
  \includegraphics[width=0.9\linewidth]{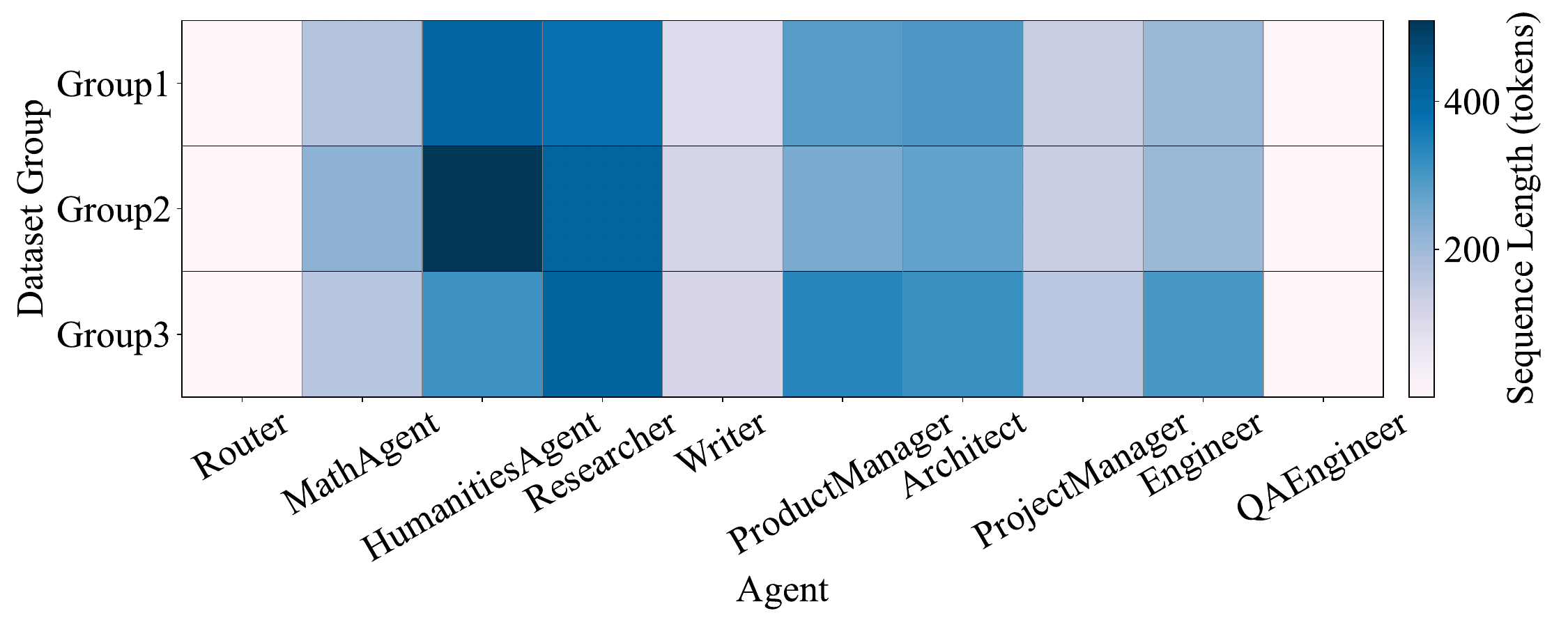}
    \caption{Average output lengths of agents from QA, RG, and CG applications across Group 1–3 datasets.}
  \Description{Heatmap showing the average output lengths of agents from QA, RG, and CG applications across Group 1, Group 2, and Group 3 datasets.}
  \label{fig:tokens_heatmap}
  \vspace{-3mm}
\end{figure}

\begin{figure}
  \centering
  \includegraphics[width=0.9\linewidth]{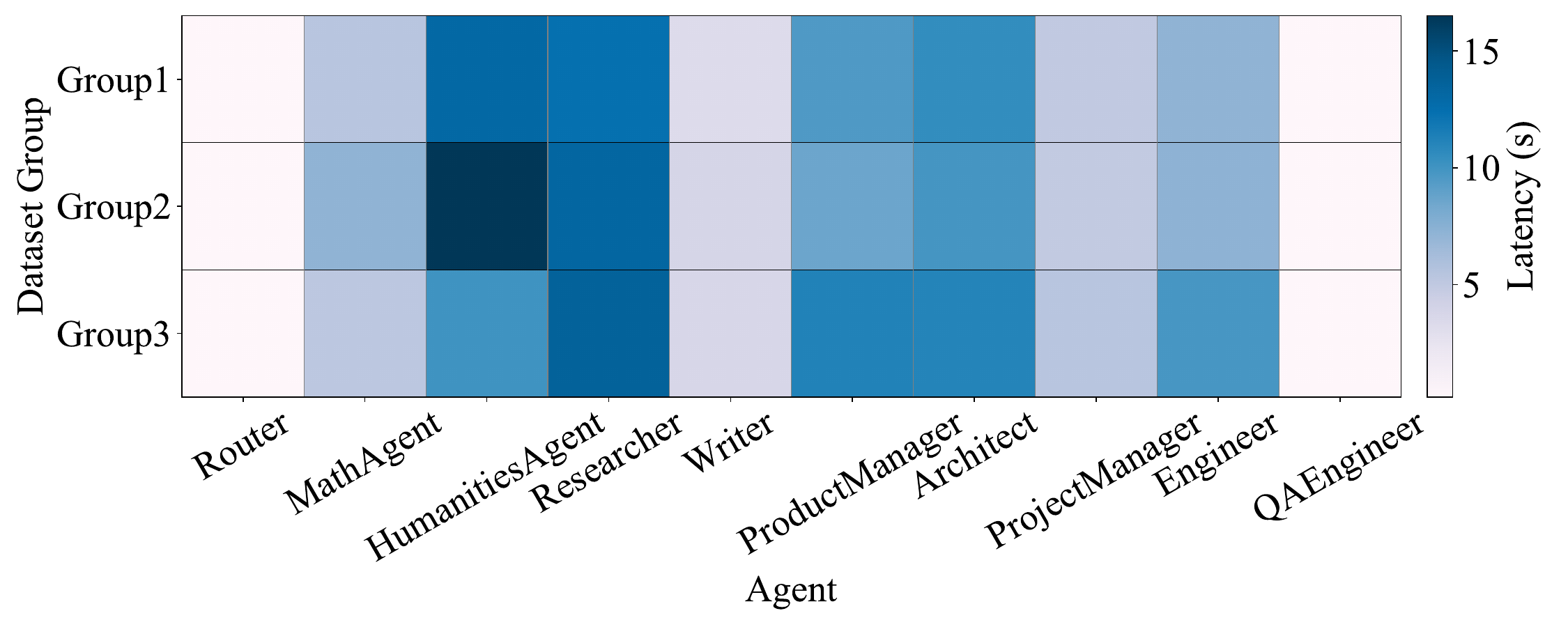}
  \caption{Average LLM inference latency of agents from QA, RG, and CG applications across Group 1–3 datasets.}
  \Description{Heatmap showing the average inference latency of agents from QA, RG, and CG applications across Group 1, Group 2, and Group 3 datasets.}
  \label{fig:latency_heatmap}
  \vspace{-3mm}
\end{figure}

We conduct evaluations on multiple datasets to further support the above observation. 
For cross-application comparison, datasets are grouped as follows: \textbf{Group 1} (G+M for QA, TQ for RG, HE for CG), \textbf{Group 2} (M+W for QA, NCD for RG, MBPP for CG), and \textbf{Group 3} (S+S for QA, NQ for RG, APPS for CG).
Figure~\ref{fig:tokens_heatmap} shows average output lengths of agents across the three groups. Agents differ significantly within each group, confirming that functional roles shape generation behavior.
Despite task variations, each agent’s behavior remains consistent across groups.
Figure~\ref{fig:latency_heatmap} shows the corresponding average inference latency, which also varies across agents but stays stable for the same agent across datasets.

From the above analysis, we can conclude that both the output length and inference latency of agents' LLM requests are affected by their functional roles, while each agent maintains stable behavior across different inputs.

\subsection{Inefficiency of Current Methods}
\subsubsection{Investigation Setup}
We utilize the state-of-the-art multi-agent orchestration systems Parrot~\cite{10.5555/3691938.3691988} and Ayo~\cite{10.1145/3676641.3716278} for investigations.
At the scheduling layer, Parrot adopts the First-Come-First-Served (FCFS) policy, while Ayo prioritizes requests with fewer remaining stages based on the stage depth in the workflow topology.
At the dispatching layer, both systems employ a Round-Robin strategy to dispatch requests across multiple LLM instances.

We use co-located workload (with datasets G+M for QA, TQ for RG, and HE for CG) and the Llama3-8B model for evaluations.
Table~\ref{tab:experimental_setup} shows our experimental configurations.
The request arrival times are derived from a popular real-world LLM inference trace~\cite{10609649}, scaled proportionally to simulate the real inter-arrival distribution.

\begin{table}
    \centering
    \scriptsize
    \setlength\tabcolsep{2pt}
    \caption{Experimental Setup}
    \label{tab:experimental_setup}
    \begin{tabular}{>{\centering\arraybackslash}m{0.28\linewidth} | >{\centering\arraybackslash}m{0.62\linewidth}}
        \hline
        \multicolumn{1}{c|}{} & \multicolumn{1}{c}{\textbf{Specifications}} \\
        \hline
        \textbf{Hardware} & \makecell[c]{NVIDIA A40 GPU (48GB global memory) $\times$ 4} \\
        \hline
        \textbf{Software} & Ubuntu 20.04 with CUDA 12.8 \\
        \hline
        \textbf{Benchmarks} & \makecell[c]{Question Answer\\ Report Generate \\ Code Generate} \\
        \hline
        \textbf{LLMs} & \makecell[c]{Llama3-8B and Llama2-13B} \\
        \hline
    \end{tabular}
    \vspace{-3mm}
\end{table}

\subsubsection{Inappropriate Request Priority Scheduling}
\begin{figure}
  \centering
  \includegraphics[width=\linewidth]{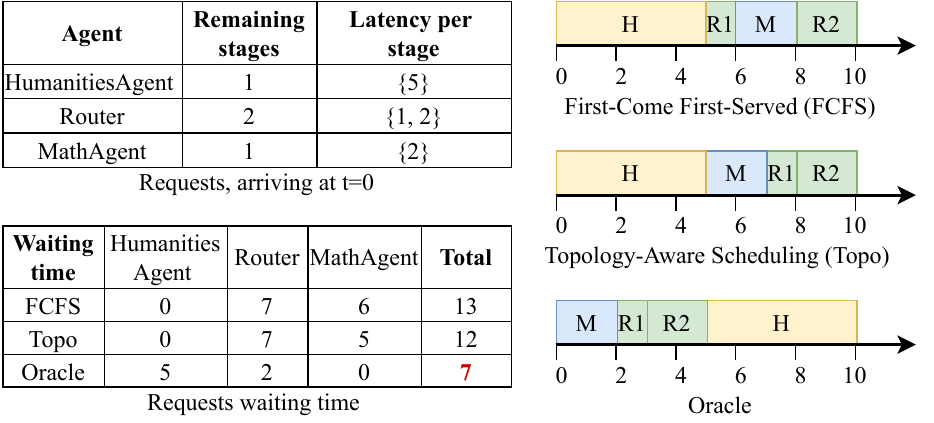}
    \caption{Examples of queuing time under FCFS, Topology-Aware (Topo), and Oracle scheduling strategies.}
    \Description{Diagram illustrating request scheduling under three strategies: FCFS, Topology-Aware, and Oracle.}
  \label{fig:moti_sort_demo}
  \vspace{-3mm}
\end{figure}

In multi-agent applications, requests from different agents often exhibit substantial variation in their remaining execution time within the overall workflow. Ideally, requests with shorter remaining execution time should be prioritized to minimize overall queuing delay. However, existing scheduling strategies fail to effectively capture such differences, resulting in mismatches between the scheduling order and the actual latency of requests.

Figure~\ref{fig:moti_sort_demo} presents an example of queuing time under three scheduling strategies. The FCFS policy fails to distinguish between requests with different latencies, causing longer requests to block shorter ones and resulting in a queuing delay of up to 13 units. Topology-Aware (Topo) scheduling prioritizes requests in later execution stages based on workflow depth, partially alleviating head-of-line blocking and reducing the queuing delay to 12 units. The ideal Oracle scheduler, informed by accurate knowledge of each request’s remaining latency, achieves the lowest queuing delay of 7 units.

\begin{figure}
  \centering
  \includegraphics[width=\linewidth]{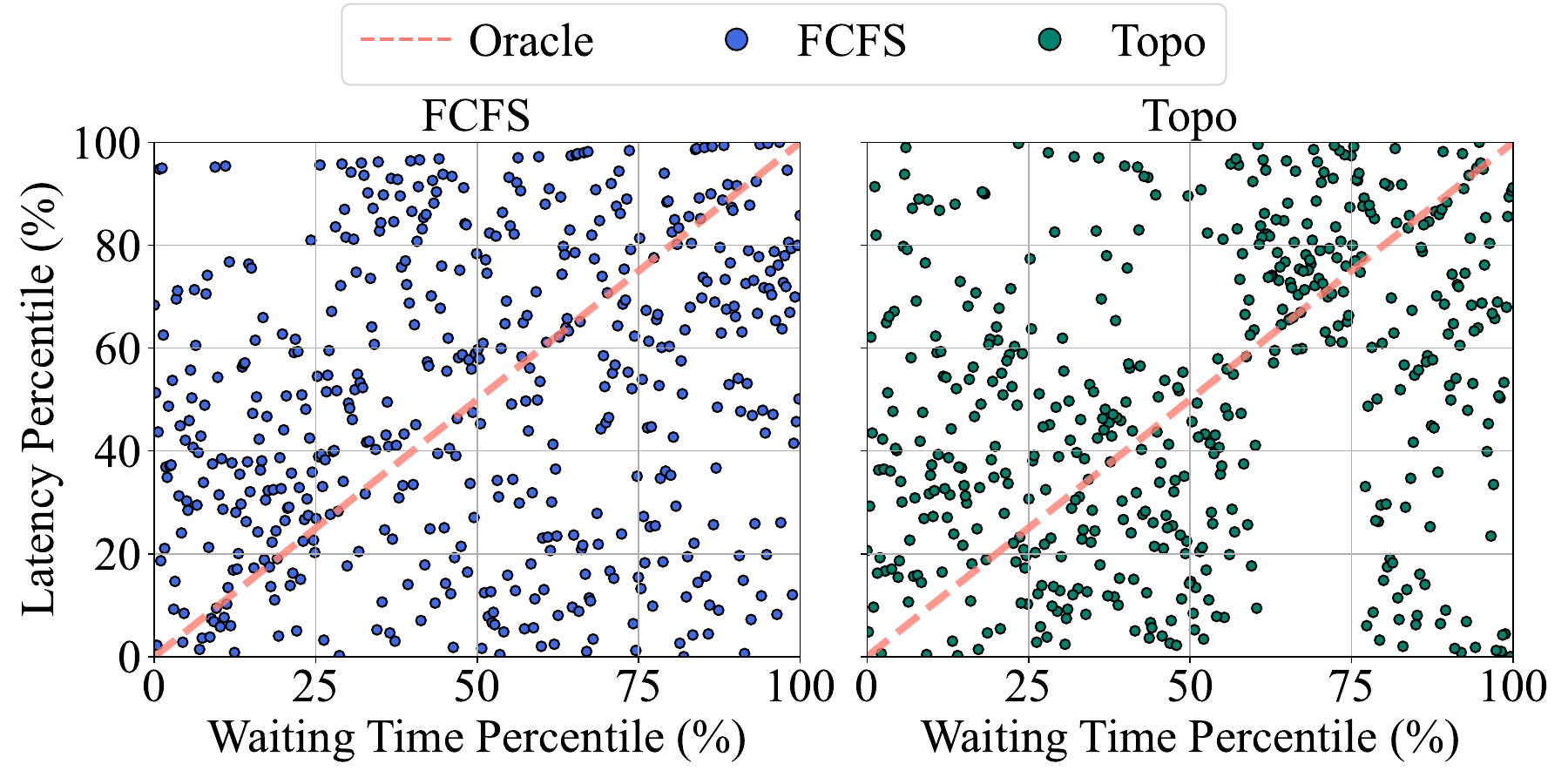}
    \caption{Comparison of request ranking by queuing time and inference latency under FCFS and Topo scheduling.}

    \Description{Comparison of request ranking by queuing time and inference latency under FCFS and Topology-Aware scheduling.}
  
  \label{fig:moti_sort_rank}
  \vspace{-6mm}
\end{figure}

Figure~\ref{fig:moti_sort_rank} shows the scatter distribution between the relative rankings of requests in queuing time and inference latency under FCFS and Topo scheduling strategies at a request rate of 8 req/s. The x-axis represents the ranking of requests based on queuing time (with points further to the left indicating earlier scheduling), while the y-axis represents the ranking of requests based on inference latency (with points lower down indicating shorter latency). Ideally, requests with lower latency should be prioritized, resulting in scatter points distributed along the diagonal.
However, this figure shows no obvious correlations between the waiting time and inference latency.

\textit{The above results indicate that effective priority scheduling is needed to properly identify differences in the remaining latency among agent requests in multi-agent applications.}

\subsubsection{Recomputations under Inefficient Request Dispatching}
Most of the current works employed a Round-Robin strategy to dispatch requests across multiple LLM instances. This overlooks the differences in memory demands among requests from different agents, leading to inefficient memory allocation and thus impacting overall performance.

\begin{figure}
  \centering
  \includegraphics[width=0.8\linewidth]{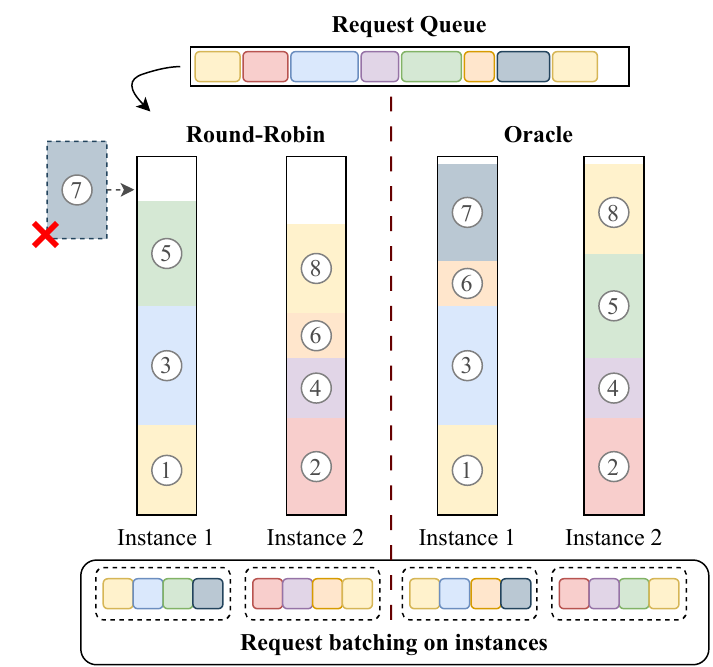}
    \caption{Request dispatching and batching across two LLM instances under Round-Robin and Oracle strategies.}

    \Description{Request dispatching and batching across two LLM instances under Round Robin and Oracle strategies.}
  
  \label{fig:moti_batching}
  \vspace{-3mm}
\end{figure}

As shown in Figure~\ref{fig:moti_batching}, requests from different agents are distributed to two LLM instances (Instances 1 and 2), and are batched within each instance, where the height of each request block represents its memory demand. Under the Round-Robin strategy, requests are assigned to instances in order of arrival. 
Because this approach does not consider the memory demand of requests and the resource status of instances, 
it causes \textit{req-7} to be preempted due to insufficient available memory, leading to resource waste. 
Under the request rate of 8 req/s, 
our results show that 18.4\% of requests are preempted during execution, leading to 14.2\% of memory resources being wasted. 

In contrast, the Oracle strategy is aware of both the memory demand of requests and the current memory usage of instances. It dispatches requests to the instance with the smallest expected peak memory usage by jointly considering the request's memory demand and the instance's current memory usage, thereby constructing more compact request batches. This effectively avoids preemption caused by memory shortages and improves overall resource utilization.

\textit{The above results indicate that the Round-Robin strategy fails to recognize differences in memory demand, thereby leading to inefficient resource usage and degrading performance.}

\subsection{Design Requirements for Efficient Multi-Agent Application Deployment}
To effectively alleviate end-to-end performance bottlenecks, three key design requirements must be addressed.

\textbf{Automated application workflow analysis.}
To achieve end-to-end aware scheduling, global information about agent requests throughout the entire application workflow needs to be obtained to perceive their remaining execution latency. 
Thus, a lightweight and automated workflow analysis mechanism needs to be designed to support efficient scheduling.

\textbf{Request priority decision based on differentiated latency distributions among agents.}
Due to the auto-regressive nature of LLM inference, the execution latency of a single request is difficult to predict precisely. Nevertheless, we observe significant differences in the latency distributions of requests from different agents. Therefore, the latency distribution characteristics of agent requests need to be leveraged to drive priority scheduling, mitigating head-of-line blocking and improving end-to-end performance.

\textbf{Request Dispatching considering memory demand differences.}
Although the memory usage of individual requests is difficult to predict accurately, significant differences are observed in the output length distributions of requests from different agents, and the input prompt length is available at dispatching. Thus, memory usage during inference needs to be dynamically perceived by combining prompt length with the output length distribution of agent requests. Requests can be dispatched accordingly to reduce preemption and improve resource utilization efficiency.
\section{Methodology}
We propose \system{}, a scalable multi-agent orchestration system that systematically optimizes the end-to-end performance of multi-agent applications. \system{} supports efficient workflow development, integrates pluggable communication mechanisms, and employs a multi-threaded architecture to handle high-concurrency requests. This enables large-scale distributed deployment of multi-agent applications with strong flexibility and scalability.

\begin{figure}
  \centering
  \includegraphics[width=\linewidth]{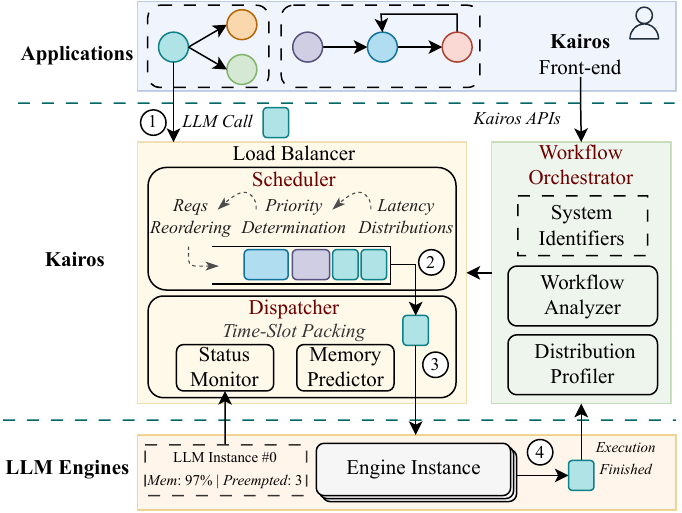}
    \caption{\system{} system overview.}
  \label{fig:overview}
  \vspace{-3mm}
\end{figure}

Figure~\ref{fig:overview} illustrates the overview of \system{}. By introducing system identifiers (\S\ref{sec:System Identifier}), \system{} supports online analysis of various dynamic workflows (\S\ref{sec:Workflow Analysis}) without requiring prior knowledge of application workflows, and continuously collects latency distributions of requests from each agent (\S\ref{sec:Latency Distribution}). Building on this foundation, \system{} designs a workflow-aware priority scheduler that dynamically prioritizes requests in the queue by leveraging workflow structure and historical latency distributions, thereby reducing overall queuing delay (\S\ref{sec:Scheduling}). For load balancing across multiple LLM instances, \system{} further proposes a memory-aware time-slot dispatcher, which intelligently dispatches requests based on the memory demands of requests and the memory status of each LLM engine to optimize end-to-end performance (\S\ref{sec:Packing}).

The Workflow Orchestrator faces the significant challenge of analyzing dynamic, complex, and unpredictable multi-agent application workflows whose structures are impossible to know beforehand. Consequently, online tracking and parsing are essential. The difficulty lies in providing developers with a lightweight solution that automatically submits the necessary contextual information for analysis at runtime, and in designing robust parsing algorithms capable of handling diverse workflow structures. \system{} overcomes this by designing system identifiers as the contextual information, and a framework that automatically propagates these identifiers at runtime. This enables online workflow analysis and the continuous collection of latency distributions.

For the Workflow-Aware Priority Scheduler, the core difficulty lies in effectively prioritizing requests to minimize end-to-end queuing delay. This is particularly challenging as individual LLM inference latency is inherently unpredictable, which leads to the inability to effectively rank requests. \system{} addresses this by leveraging the statistical diversity in remaining execution latencies across different agents. Based on the collected remaining execution latency distributions for each agent, \system{} designs an agent-level priority determination mechanism and performs priority scheduling accordingly. An intra-agent scheduling mechanism complements this by prioritizing requests with longer cumulative latency to reduce tail latency.

Finally, the Memory-Aware Time-Slot Dispatcher is challenged by efficiently dispatching diverse agent requests to LLM instances to minimize preemption and optimize GPU memory utilization under dynamic memory demands. \system{} solves this by modeling the dynamic memory usage of each request and designing a time-slot packing strategy. This involves discretizing the future timeline to evaluate expected memory usage across instances and selecting the instance with the lowest expected total peak memory usage.

The overall workflow of \system{} is as follows.
\ding{192} LLM requests from various agents are submitted to \system{}'s Load Balancer and enqueued into the request queue.
\ding{193} Based on real-time analysis from the Workflow Orchestrator, the Workflow-Aware Priority Scheduler evaluates the priority of requests from different agents and schedules the highest-priority request from the queue.
\ding{194} For the dequeued request, the Memory-Aware Time-Slot Dispatcher further selects the optimal LLM instance. Specifically, the dispatcher leverages the Status Monitor to track real-time memory usage and the number of preempted requests on each instance. Together with the analysis from the Workflow Orchestrator, the Memory Predictor uses this information to determine the most suitable instance for execution.
\ding{195} Once a request is completed, the Workflow Orchestrator collects its execution information and incrementally updates the Workflow Analyzer and the Distribution Profiler.

\system{} is implemented in about 6,000 lines of Python code. It seamlessly integrates with vLLM~\cite{kwon2023efficient} as the underlying LLM inference engine via its standard APIs (for request execution and status monitoring) and leverages Kafka~\cite{kafka} for efficient inter-agent communication. \system{} adopts a distributed deployment architecture, deploying Workflow Orchestrator, Load Balancer, and agent processes via multi-processing, and using multi-threading to handle high-concurrency requests. \system{} provides an HTTP interface compatible with the OpenAI API format~\cite{openai_api} to serve multi-agent applications.

\section{Workflow Orchestrator}
In this section, we detail \system{}'s workflow orchestration capabilities, covering its system identifiers, automated workflow analysis, and continuous latency distribution analysis.
\subsection{System Identifier}
\label{sec:System Identifier}
To support workflow analysis and historical data collection, \system{} introduces a set of system identifiers.

\begin{itemize}
    \item \textbf{Agent Name:} Used to distinguish different agents, reflecting their independent behavioral patterns and execution characteristics.
    \item \textbf{Message ID:} Each user request carries a globally unique \texttt{msg\_id} to track all agent requests involved in the entire workflow.
    \item \textbf{Upstream Name:} Records the name of the immediate preceding agent that triggered the current request in the workflow, to obtain the upstream-downstream relationships for reconstructing the workflow.
    \item \textbf{Execution Timestamps:} Records timestamps for LLM execution's start and completion for each task, enabling measurement of task spans and temporal causality analysis for workflow reconstruction.
\end{itemize}

The use of system identifiers is almost transparent to developers. The \textbf{Agent Name} only needs to be explicitly specified by the user during the workflow definition phase, as shown in Listing~\ref{lst:kairos_api}. The \textbf{Message ID}, \textbf{Upstream Name}, and \textbf{Execution Timestamps} are automatically generated by \system{} and transparently propagated across agents through the communication mechanism.

\begin{lstlisting}[style=mystyle, language=Python, caption={Usage Example of the Kairos API}, label={lst:kairos_api}, float]
from Kairos import BaseAgent, Workflow
### Implementing the Agent
PROMPT = "You're a router assistant, ..."
# Subclass BaseAgent and override _run_impl
class Router(BaseAgent):
    def _run_impl(self, input_data, metadata):
        # Fill the prompt
        question = input_data.get("question")
        prompt = PROMPT.format(question=question)
        # Use the provided function
        # to request the LLM service
        result = self.generate(prompt, metadata)
        # Determine downstream agent from result
        next_agent = ...
        # Forward result to the downstream agent
        return {"question": question}, next_agent
### Defining the workflow
workflow = Workflow()
workflow.add_engine("vllm", model="Llama-3-8B")
workflow.add_agent(agent_name="Router", agent_class=Router, use_model="Llama-3-8B")
\end{lstlisting}

\subsection{Automated Workflow Analysis}
\label{sec:Workflow Analysis}
\begin{figure}
  \centering

  \subcaptionbox{Parallel workflow\label{fig:parallel_workflow}}{%
    \includegraphics[width=0.45\linewidth]{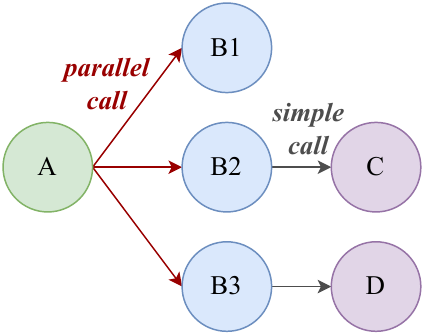}}
    \subcaptionbox{Parallel task span analysis\label{fig:parallel_span}}{%
    \includegraphics[width=0.49\linewidth]{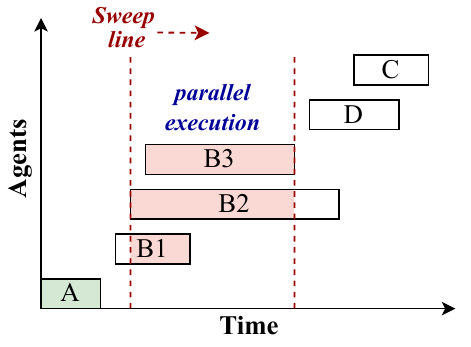}}%
    
    \par\vspace{0.5em}
    
  \subcaptionbox{Sequential workflow\label{fig:sequential_workflow}}{%
    \includegraphics[width=0.45\linewidth]{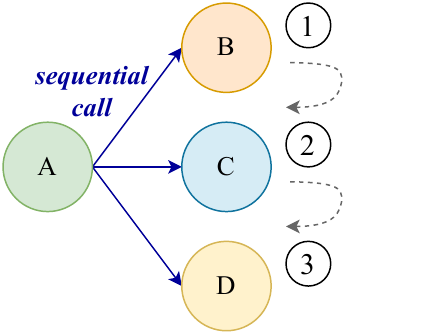}}
    \subcaptionbox{Sequential task span analysis\label{fig:sequential_span}}{%
    \includegraphics[width=0.49\linewidth]{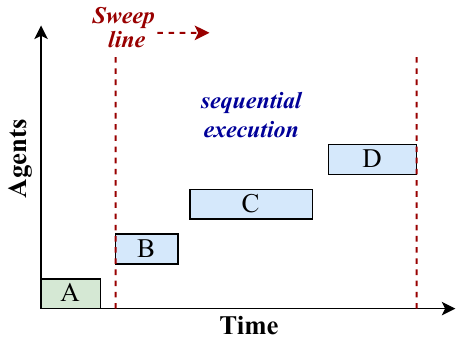}}%

  \caption{Examples of complex workflows and corresponding runtime analysis: Multiple downstream parallel and sequential execution patterns.}
  
  \label{fig:complex_workflow}
  \vspace{-6mm}
\end{figure}

We observe that beyond typical workflows as shown in Figure~\ref{fig:benchmark}, more complex workflow structures also exist~\cite{wang2024moa,qiao2024taskweaver,zhang2024ufoui}, as illustrated by Figure~\ref{fig:complex_workflow}. 
To enable comprehensive and adaptable support for these intricate multi-agent applications, \system{} introduces an automated workflow analysis mechanism.
Unlike existing methods that require explicitly predefined workflow definitions or data dependencies for their analysis, this mechanism can automatically reconstruct the complete call graph at runtime based on temporal and upstream-downstream causal relationships.

Specifically, \system{} leverages the \textbf{Message ID} to identify and collect all request data belonging to the same workflow, including the \textbf{Agent Name}, \textbf{Upstream Name}, and \textbf{Execution Timestamps}, for generating upstream-downstream relationships and temporal task span information.

Both \textbf{Upstream Name} and \textbf{Execution Timestamps} are critical for robust workflow reconstruction. For instance, relying solely on the \textbf{Upstream Name} would only reveal direct calling relationships, but it would be insufficient to distinguish whether these multiple downstream calls are executed in parallel (Figure~\ref{fig:parallel_workflow}) or sequentially (Figure~\ref{fig:sequential_workflow}).
Conversely, using only \textbf{Execution Timestamps} could allow for chronological ordering, but might erroneously reconstruct upstream-downstream dependencies. For example, the multiple downstream sequential workflow (Figure~\ref{fig:sequential_workflow}) might be erroneously interpreted as a sequential chain (A $\to$ B $\to$ C $\to$ D), rather than correctly identifying Agent A as the sole sequential initiator. Such misinterpretations would lead to incorrect dependency analysis and distorted workflow topologies.

Based on the upstream-downstream dependencies, \system{} first constructs a preliminary workflow execution graph. Subsequently, \system{} analyzes the call patterns (parallel or sequential) of each node's outgoing edges within this graph. If a node has only a single downstream dependency, that edge represents a simple call. 
For nodes with multiple downstream dependencies, \system{} utilizes a sweep-line algorithm, leveraging temporal span information to identify parallel execution requests, as illustrated in Figure~\ref{fig:parallel_span} and Figure~\ref{fig:sequential_span}.

Through this mechanism, \system{} can reconstruct diverse multi-agent application workflows online, including typical structures in Figure~\ref{fig:benchmark}, providing a structural and semantic foundation for data collection and scheduling optimization.

\subsection{Latency Distribution Analysis}
\label{sec:Latency Distribution}
\system{} continuously collects LLM execution samples from various agents at runtime and statistically analyzes their execution latency characteristics. Specifically, it constructs and maintains two types of distributions.

\textbf{1. Single-Request Execution Latency Distribution.} \system{} continuously collects execution latency samples of LLM requests from each agent online and employs an exponentially increasing sampling strategy to determine distribution convergence. 
Each time the sample size doubles, \system{} computes the Wasserstein distance~\cite{wasserstein} between the current and previous distributions. This distance captures both the shape and positional shift between distributions. If this distance falls below a predefined threshold, the distribution is considered converged and can be utilized for subsequent optimization. 
This distribution guides the request packing strategy described in \S\ref{sec:Packing}.

\textbf{2. Remaining Execution Latency Distribution.} \system{} further constructs the remaining execution latency distribution for each agent, which reflects the time required to complete the end-to-end workflow from the current execution stage. Based on the workflow structure extracted in \S\ref{sec:Workflow Analysis}, \system{} computes the remaining end-to-end latency for each request. These per-request measurements are then aggregated to construct the remaining execution latency distribution for each agent.
This distribution guides the request scheduling strategy described in \S\ref{sec:Scheduling}.

In particular, for applications with autonomous planning capabilities (Figure~\ref{fig:benchmark-qa}), some agents may have multiple downstream execution paths. In such cases, \system{} collects the remaining execution latency samples separately for each path and merges them into the overall distribution. This is based on the intuition that historical data reflects the actual distribution of user tasks and can approximate the likelihood of future execution paths. Thus, paths with higher historical frequency (e.g., \texttt{Router} to \texttt{MathAgent} for frequent math questions) contribute proportionally more to the remaining latency distribution of the upstream agent, thereby reflecting anticipated future user request patterns.

\section{Workflow-Aware Priority Scheduler}
\begin{figure}
  \centering
  \includegraphics[width=0.97\linewidth]{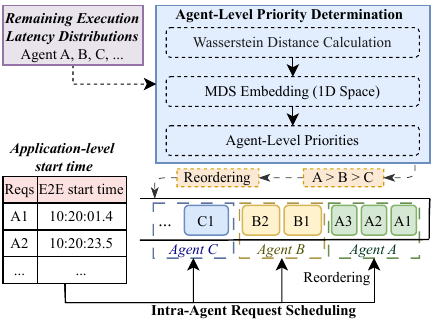}
    \caption{Workflow-aware priority scheduling strategy.}
  \label{fig:workflow_aware_priority_scheduling_strategy}
  \vspace{-6mm}
\end{figure}

\label{sec:Scheduling}
As previously described, \system{} continuously constructs the remaining execution latency distribution for each agent. Based on this distribution, we introduce a novel \textbf{workflow-aware priority scheduling strategy}. As illustrated in Figure~\ref{fig:workflow_aware_priority_scheduling_strategy}, it first determines agent-level priorities to order agents and then sorts intra-agent requests through application-level start time, aiming to reduce the overall queuing latency.
The intuition is that while an individual request's remaining execution latency is uncertain, the relative latency across agents remains statistically stable. Prioritizing agent requests with shorter remaining latency allows \system{} to globally optimize the queuing order, reducing end-to-end latency.

\subsection{Agent-Level Priority Determination Mechanism}
\label{sec:agent-level_mechanism}
\system{} introduces an agent-level priority determination mechanism that leverages differences in \textbf{remaining execution latency distributions}.

\system{} first measures the distributional differences among agents using the Wasserstein distance~\cite{wasserstein}. This provides a robust and intuitive basis for comparing the remaining execution latencies across different agents. A pairwise distance matrix is then constructed based on these Wasserstein distances. To enable comparison of the relative latency among all agents in a unified dimension, \system{} applies Multidimensional Scaling (MDS)~\cite{mds} to embed the distance matrix into a one-dimensional coordinate space. MDS preserves the original distance relationships as much as possible while transforming the pairwise distributional differences into an interpretable 1D representation, facilitating subsequent priority determination.

However, the coordinate space obtained by MDS is directionless and only reflects the relative differences among agents, making it impossible to directly determine which agents have shorter remaining execution latency. To assign a priority direction to the coordinate space, \system{} introduces an ideal “zero latency” distribution as an anchor point.

\system{} includes the “zero latency” distribution into the set of all agents' distributions, computes the Wasserstein distances among all distributions, and maps these distributions into a 1D coordinate space via MDS. Agents positioned closer to the ideal distribution anchor in this coordinate space have original distributions more similar to the “zero latency” ideal distribution, indicating shorter remaining execution latency and thus are assigned higher scheduling priorities.

Through this process, \system{} constructs a stable priority determination mechanism without requiring precise prediction of individual request execution latency, thereby effectively supporting the priority scheduling.

\subsection{Intra-Agent Request Scheduling Mechanism}
As discussed in Section~\ref{sec:agent-level_mechanism}, \system{} determines inter-agent request priorities. This section further explores the intra-agent request scheduling mechanism.

Since intra-agent requests originate from the same application, they typically share the same end-to-end latency constraints. However, due to the inherent uncertainty of autoregressive inference in LLMs, these requests may accumulate significantly different latencies in prior LLM inference stages. Some have already incurred shorter delays, while others have incurred longer cumulative latency. To further reduce end-to-end tail latency, \system{} prioritizes requests that have already experienced longer delays in prior stages.

Specifically, as described in Section~\ref{sec:System Identifier}, \system{} assigns a globally unique Message ID to each request upon its arrival at the frontend. Based on this identifier, \system{} records the request’s \textbf{application-level start time} (i.e., its arrival time at the frontend), which is used to measure the accumulated latency in prior stages. \system{} maintains a global mapping \texttt{\{msg\_id: e2e\_start\_time\}} to support intra-agent request scheduling decisions.
At scheduling, \system{} uses this start time as the ordering criterion for intra-agent requests, prioritizing those with earlier application-level start time. This strategy differs from the default scheduling strategy of current LLM engines that relies solely on the request-level start time (i.e., the arrival time at the current agent stage), as it captures the accumulated latency of requests in prior stages, thereby effectively reducing end-to-end tail latency.

\section{Memory-Aware Time-Slot Dispatcher}
\label{sec:Packing}
After determining the scheduling order, \system{} must decide which LLM instance each request should be dispatched to.
As discussed in Section~\ref{sec:motivation_character}, requests from different agents exhibit significant variations in memory usage.
To address this, we design a \textbf{memory-aware time-slot packing strategy} that matches requests to instances in a way that balances memory load. For example, requests with higher memory demands can be preferentially dispatched to instances with larger available memory, thereby reducing the risk of out-of-memory (OOM) failures and improving resource utilization.

However, memory usage varies dynamically over time, making static matching strategies insufficient to capture the actual resource demands of requests and the real-time status of instances. Therefore, we model the memory usage of each request as a linear function over time to assist subsequent packing decisions.
The KV Cache usage of request $i$ during its execution can be approximated by Equation~\ref{equa:memory_usage}, where $P_i$ denotes the memory usage corresponding to the prompt of request $i$ (i.e., the memory usage of the prefill phase), which can be computed online by \system{} in real time; $k$ represents the decoding speed of the request, corresponding to the rate of memory usage increase, and is determined through prior hardware profiling; and $t_i,start$ and $t_i,end$ denote the start and end times of request $i$’s decoding phase, respectively.

\vspace{-4mm}
\begin{equation}
\label{equa:memory_usage}
f_i(t) = 
\begin{cases}
P_i + k \times (t - t_{i,\text{start}}), & \text{if } t_{i,\text{start}} < t < t_{i,\text{end}} \\[2pt]
0, & \text{otherwise}
\end{cases}
\end{equation}

In this equation, $t_i,start$ can be obtained by \system{} in real time, whereas $t_i,end$ is difficult to predict precisely. As discussed in Section~\ref{sec:motivation_character}, since the decoding latency accounts for the majority of the overall latency, we approximate $t_i,end$ using the expected execution time $T_i$, as shown in Equation~\ref{equa:end_time}.

\vspace{-3.5mm}
\begin{equation}
\label{equa:end_time}
t_i,end = t_i,start + T_i
\end{equation}

Based on the \textbf{single-request execution latency distribution} derived from historical data, we select the point with the highest probability density as the expected execution time of the request. This statistic reflects the most common execution latency experienced by requests from the agent, thereby closely representing actual execution behavior.

Building on the above single-request model, we can model the memory usage during request batching as shown in Equation~\ref{equa:batching}, where $\mathcal{A}_j$ denotes the set of requests currently assigned to instance $j$, and $f_i(t)$ represents the memory usage function of request $i$. The function $F_j(t)$ characterizes the memory usage of instance $j$ over future time, serving as the foundation for our subsequent packing strategy.

\vspace{-3.5mm}
\begin{equation}
\label{equa:batching}
F_j(t) = \sum_{i \in \mathcal{A}_j} f_i(t)
\end{equation}

\begin{figure}
  \centering
  \includegraphics[width=0.75\linewidth]{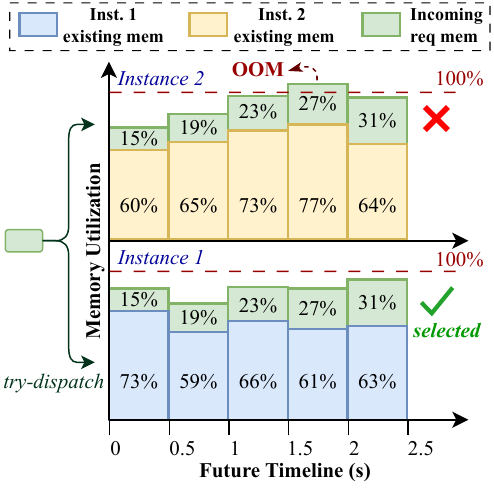}
    \caption{Memory-aware time-slot packing strategy.}
  \label{fig:memory_aware_time_slot_packing_strategy}
  \vspace{-6mm}
\end{figure}

We design a time-slot packing strategy that determines the allocation target for each request. To reduce decision-making overhead, we divide the future timeline into several fixed-length time slots, allowing for fast evaluations of instance memory usage based on these discrete slots. The key process is illustrated in Figure~\ref{fig:memory_aware_time_slot_packing_strategy} and involves two steps:

\textbf{1. Time Slot Partitioning and Memory Usage Accumulation.}
Shorter slot lengths enable finer-grained evaluation but incur higher computational overhead. Through empirical observations, we find that a slot length of 0.5 seconds offers a favorable trade-off. When dispatching a request, \system{} first identifies the set of time slots $\mathcal{S}$ it will span. Then, \system{} computes the memory usage of the request in each time slot and accumulates it with the memory usage of existing requests already assigned to the instance for the corresponding slots, thus obtaining the total expected memory usage. Finally, \system{} scans through all time slots in $\mathcal{S}$. If the memory usage in any slot exceeds the total memory capacity of the instance, the instance is considered temporarily unavailable for dispatch.

\textbf{2. Instance Selection.} For each request, \system{} evaluates the availability of all LLM instances in parallel to accelerate the process. If none of the instances are available, the request remains in the scheduling queue, awaiting the next scheduling round. Otherwise, \system{} selects the instance with the lowest expected total peak memory usage in the future among the available instances, aiming to minimize the risk of OOM failures caused by estimation errors.

While this strategy proactively balances memory load, some special cases exist because of the unpredictable nature of LLM inference. To address this, for requests that execute faster than anticipated, \system{} immediately removes its memory usage from subsequent time slots to prevent interference with future request dispatching. Conversely, for requests that execute slower than anticipated, \system{} monitors for potential OOM events, temporarily suspending new dispatches to the affected instance upon detection.  
Compared to simple strategies that, for instance, rely on static memory thresholding (e.g., 90\% memory capacity as a threshold), which often sacrifices memory utilization, these adaptive measures can enhance system stability and resource efficiency.
\begin{figure*}[htbp]
    \centering
    \includegraphics[width=1\linewidth]{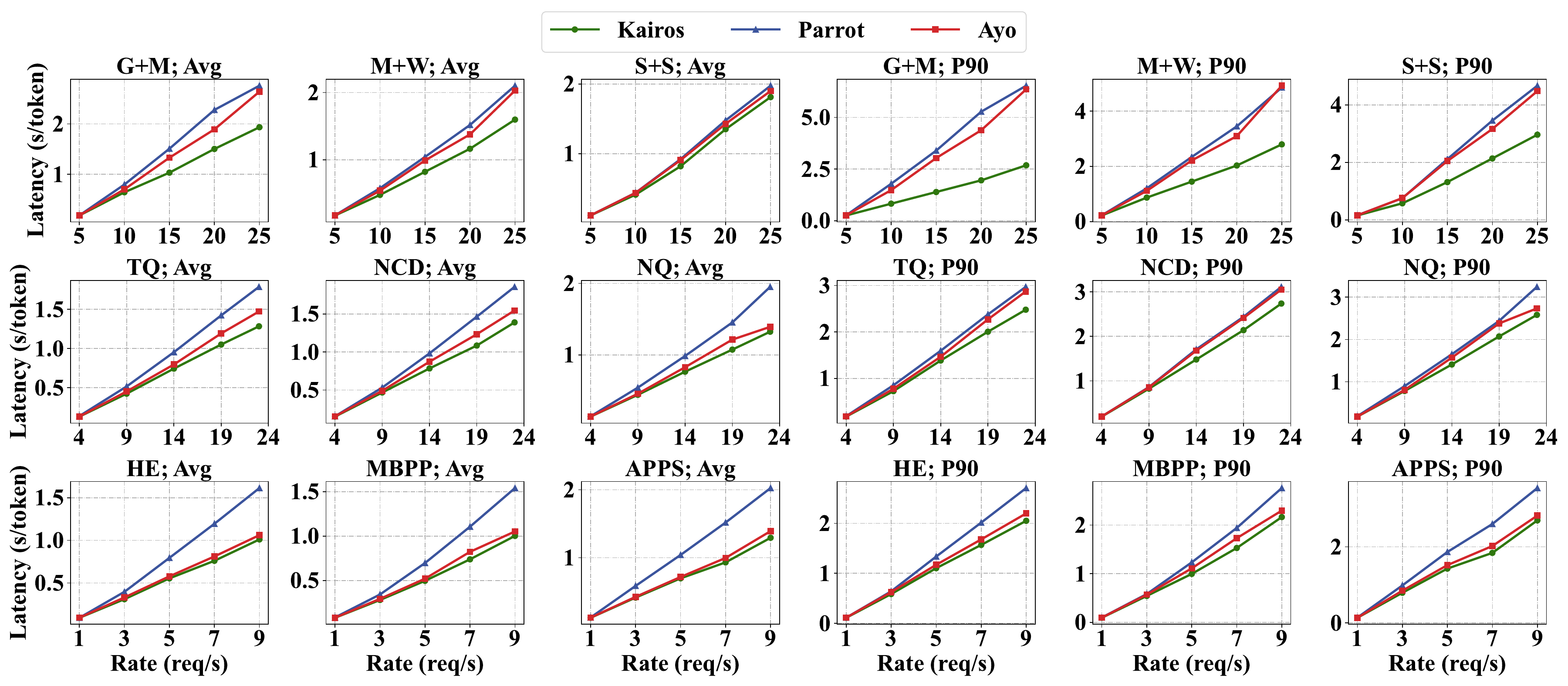}
    \caption{
    End-to-end performance of individual applications:
    \textit{Question Answer} (1st row), \textit{Report Generate} (2nd row), and \textit{Code Generate} (3rd row).
    The subcaption of each subfigure indicates the dataset and the latency type (average or tail).
    }
    \label{fig:single_app_latency}
    \vspace{-3mm}
\end{figure*}

\section{Evaluation}
In this section, we first evaluate \system{} in minimizing end-to-end latency for multi-agent applications, and then evaluate the effectiveness of \system{}'s each module.

\subsection{Experimental Setup}
\textbf{Testbed Configuration.} We conduct our experiments using the configurations detailed in Table~\ref{tab:experimental_setup}. The workloads and their corresponding datasets are described in Section~\ref{sec:motivation_description}.

\textbf{Baseline.} We select two representative state-of-the-art systems as baselines. All methods adopt the same LLM engine configuration to ensure fairness.

\textit{Parrot~\cite{10.5555/3691938.3691988}}: The First-Come-First-Serve (FCFS) scheduling strategy is adopted for scheduling requests from different agents.
Moreover, Parrot dispatches requests from different agents of the same application to multiple LLM instances in a Round-Robin manner.

\textit{Ayo~\cite{10.1145/3676641.3716278}}: A priority scheduling strategy based on workflow topology depth is introduced, prioritizing requests with fewer remaining downstream stages. Moreover, requests are dispatched in a Round-Robin manner.

\textbf{Loads.} To simulate the dynamic arrival patterns of requests in the real world, we construct the loads according to a production-level LLM inference trace~\cite{10609649}, preserving the original distributions of inter-request intervals through proportional sampling. We adjust the overall load rate so that the average queueing time ratio (i.e., the ratio of queueing time to end-to-end time) ranges from 0\% to 90\%, thereby covering a variety of scenarios from low to high loads.

\textbf{Metrics.} We adopt program-level token latency~\cite{luo2025autellix} as the performance metric to measure the end-to-end performance of multi-agent applications. This metric is defined as the end-to-end response time divided by the number of tokens generated,
with lower values indicating better performance. 
For brevity, we refer to program-level token latency as ``latency'' in the following. We report both the average and the tail latency to reflect the overall performance.

\subsection{End-to-end Performance}
\label{sec:single_e2e}

In this section, we evaluate the end-to-end performance of \system{} against baselines for individual application workloads, specifically across three representative applications and various datasets. We use Llama3-8B as the serving LLM.

As shown in Figure~\ref{fig:single_app_latency}, \system{} consistently outperforms Parrot and Ayo across all evaluated applications and datasets. Compared to Parrot, \system{} reduces both average and tail latency, with average decrease ranging from 17.8\% to 28.4\%, and P90 tail latency decrease ranging from 19.1\% to 28.6\%. Furthermore, \system{} also achieves significant gains over Ayo, decreasing average latency by 5.8\% to 10.8\%, and P90 latency by 13.4\% to 20.2\%.

The advantages stem from \system{}'s workflow-aware priority scheduling and memory-aware time-slot dispatching. 
The scheduling prioritizes agent requests based on their remaining execution latencies, allowing requests that can complete faster to execute first, thereby alleviating request-level head-of-line blocking. 
The dispatching perceives the memory demand differences of agent requests and dynamically assigns requests to suitable LLM instances, effectively reducing request preemption and improving resource efficiency.
In contrast, Parrot's FCFS fails to capture execution differences among requests.
While Ayo considers the workflow topology depth offers an improvement, it cannot differentiate requests with varied execution latencies at the same depth. 
Moreover, both of them employ the Round-Robin that disregards memory demand differences. This results in low memory utilization and frequent preemptions.

Notably, for QA on the \textit{S+S} dataset, \system{}'s advantage in average latency is less pronounced. This is because SocialIQA, a social science question dataset, leads the HumanitiesAgent to produce shorter outputs compared to historical questions, thereby reducing latency differences between agents and weakening scheduling effectiveness. For RG and CG applications, \system{}'s performance improvement over Ayo is comparatively modest. This is because both applications predominantly contain simple linear workflow structures, which allows Ayo’s priority scheduling to achieve performance close to \system{}’s. Nevertheless, \system{} still achieves these gains, primarily benefiting from its memory-aware time-slot packing strategy in these scenarios.

In summary, \system{} outperforms baselines in both average and tail latencies for individual application workloads.

\subsection{Co-located Applications}
\label{sec:co-location}

In production-level LLM services, requests from multiple applications typically share LLM instances to improve resource utilization and reduce costs~\cite{10.5555/3691938.3691988,298685}.  
To evaluate \system{}'s performance in this multi-application co-location scenario, we use co-located workload (with datasets G+M for QA, TQ for RG, and HE for CG) and the Llama3-8B model.

\begin{figure}
  \centering
  \includegraphics[width=0.8\linewidth]{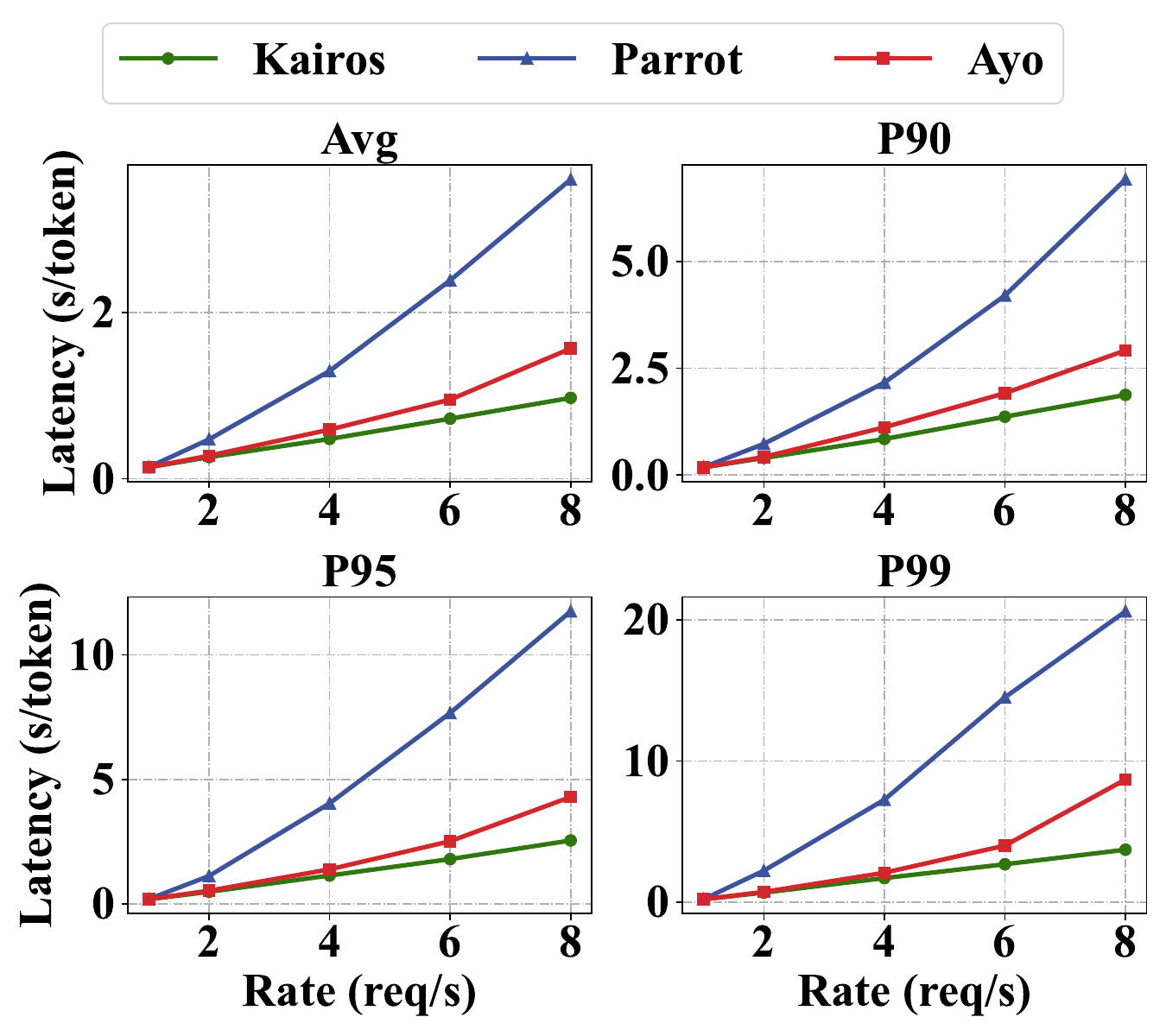}
    \caption{Latency performance of co-located applications (QA, RG, and CG) using Llama3-8B.}
  \label{fig:colocation_latency_8B}
  \vspace{-6mm}
\end{figure}

As shown in Figure~\ref{fig:colocation_latency_8B}, \system{} shows a pronounced performance advantage in this complex scenario. Compared to Parrot, \system{} decreases the average latency 
by 45.1\% to 72.8\%, P90 latency by 45.4\% to 72.9\%, P95 latency by 56.8\% to 78.3\%, and P99 latency by 69.6\% to 81.9\%. Furthermore, compared to Ayo, \system{} also decreases the average latency
by 6.1\% to 37.9\%, P90 latency by 5.6\% to 35.7\%, P95 latency by 8.4\% to 40.6\%, and P99 latency by 6.6\% to 57.2\%.

\system{}'s advantage lies in precisely identifying differences in remaining execution latency and memory demands among diverse agent requests from multiple applications. This is achieved through unified cross-application modeling, enabling fine-grained and efficient request scheduling and dispatching.
In contrast, Parrot fails to identify execution differences among requests, resulting in severe queueing delays and resource waste. Ayo lacks a cross-application joint scheduling, thus its overall performance improvement is limited.

These results demonstrate that \system{} exhibits significant advantages in the multi-application co-location scenario, which commonly exists in the real world.

\subsection{Priority Ordering Accuracy}

To reduce overall queueing latency, \system{} introduces a workflow-aware priority scheduling strategy. 
To further validate the effectiveness of this strategy, this subsection quantitatively evaluates its sorting accuracy within the queue.

\textbf{Experimental Scenarios.} We construct 10 experimental scenarios, including nine single-application scenarios from Section~\ref{sec:single_e2e}, and a scenario representing the co-located workload from Section~\ref{sec:co-location}. For comparison with the theoretically optimal sorting, each scenario uses all historical execution data to simulate requests in the queue.

\textbf{Definition of Accuracy.} 
\system{}'s scheduling strategy prioritizes requests based on agent-level priority, derived from the remaining execution latency distributions of different agents. 
To quantify its practical effectiveness, we measure request-level sorting accuracy. This is calculated by forming \textbf{request pairs} for each request in the queue, comparing that request with all other agent requests, and determining the proportion of correctly sorted pairs. The overall sorting accuracy for a scenario is the average of these request-level accuracies. A higher accuracy indicates that the system more effectively identifies remaining execution latency differences and maps them to the actual scheduling order.

\begin{figure}
  \centering
  \includegraphics[width=0.9\linewidth]{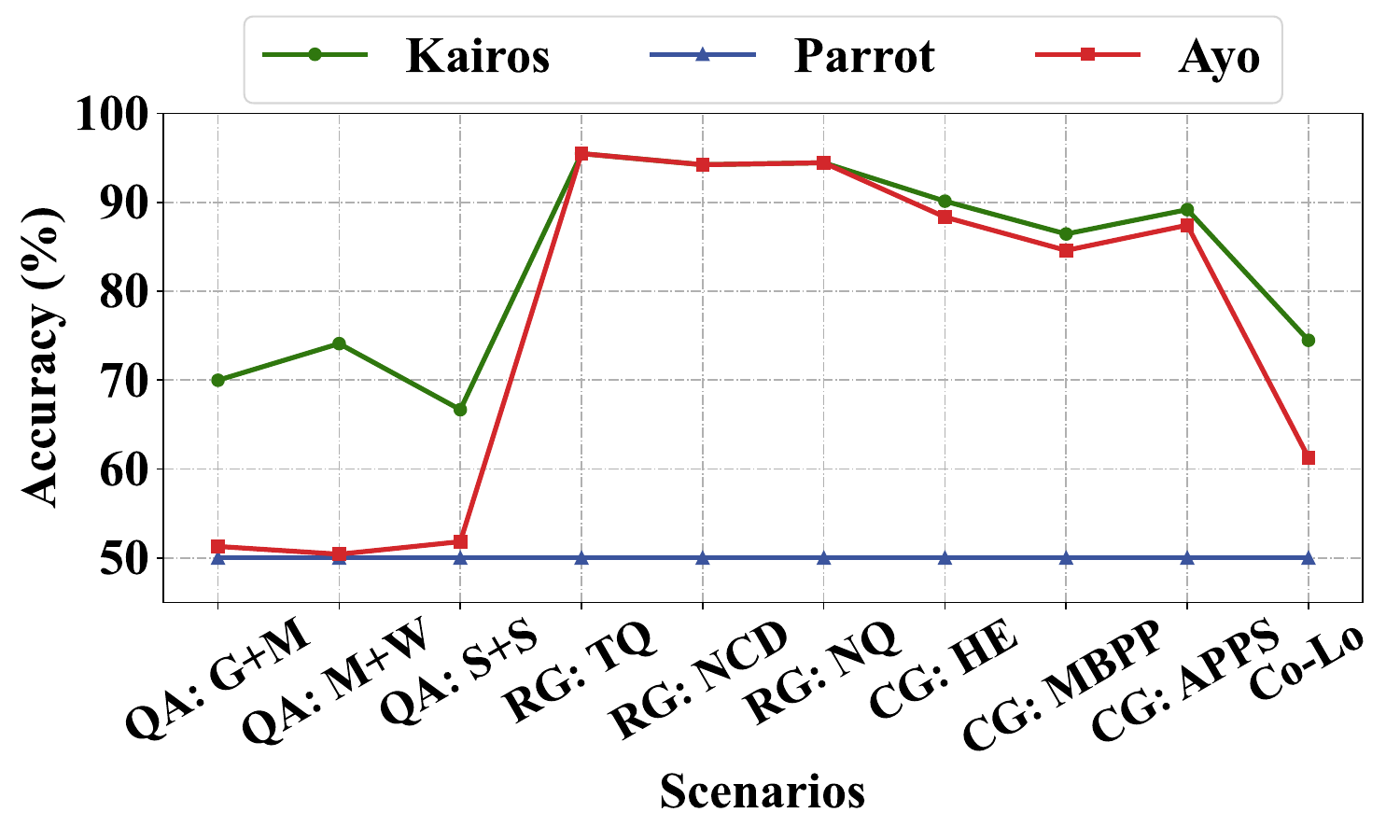}
    \caption{Sorting accuracy across 10 experimental scenarios, including single-application workloads on three datasets for each application and a co-located workload.}
  \label{fig:accuracy}
  \vspace{-3mm}
\end{figure}

\textbf{Results.} As shown in Figure~\ref{fig:accuracy}, \system{} shows superior sorting accuracy across various scenarios, achieving an average accuracy of 83.5\%, while Ayo exhibits an average accuracy of 75.9\%. 
Parrot adopts the FCFS strategy, which results in a random sorting accuracy of 50\%, because for any given request pair, either request may arrive and be scheduled first.
The better performance is attributed to that \system{} can identify remaining execution latency differences among agent requests in both single- and cross-application scenarios, and accurately map these differences to the actual scheduling. 

We can also observe specific nuances in certain scenarios: for RG and CG applications, Ayo's accuracy is close to \system{}'s because their workflows are either entirely linear (for RG) or predominantly contain simple linear structures (for CG), allowing Ayo’s priority scheduling strategy to perform comparably.

\subsection{Scalability to Larger LLM}

To evaluate \system{}'s performance with a larger LLM, we replace Llama3-8B with Llama2-13B and use the co-located workload from Section~\ref{sec:co-location} to maintain the production-level LLM service scenario. 

\begin{figure}
  \centering
  \includegraphics[width=0.85\linewidth]{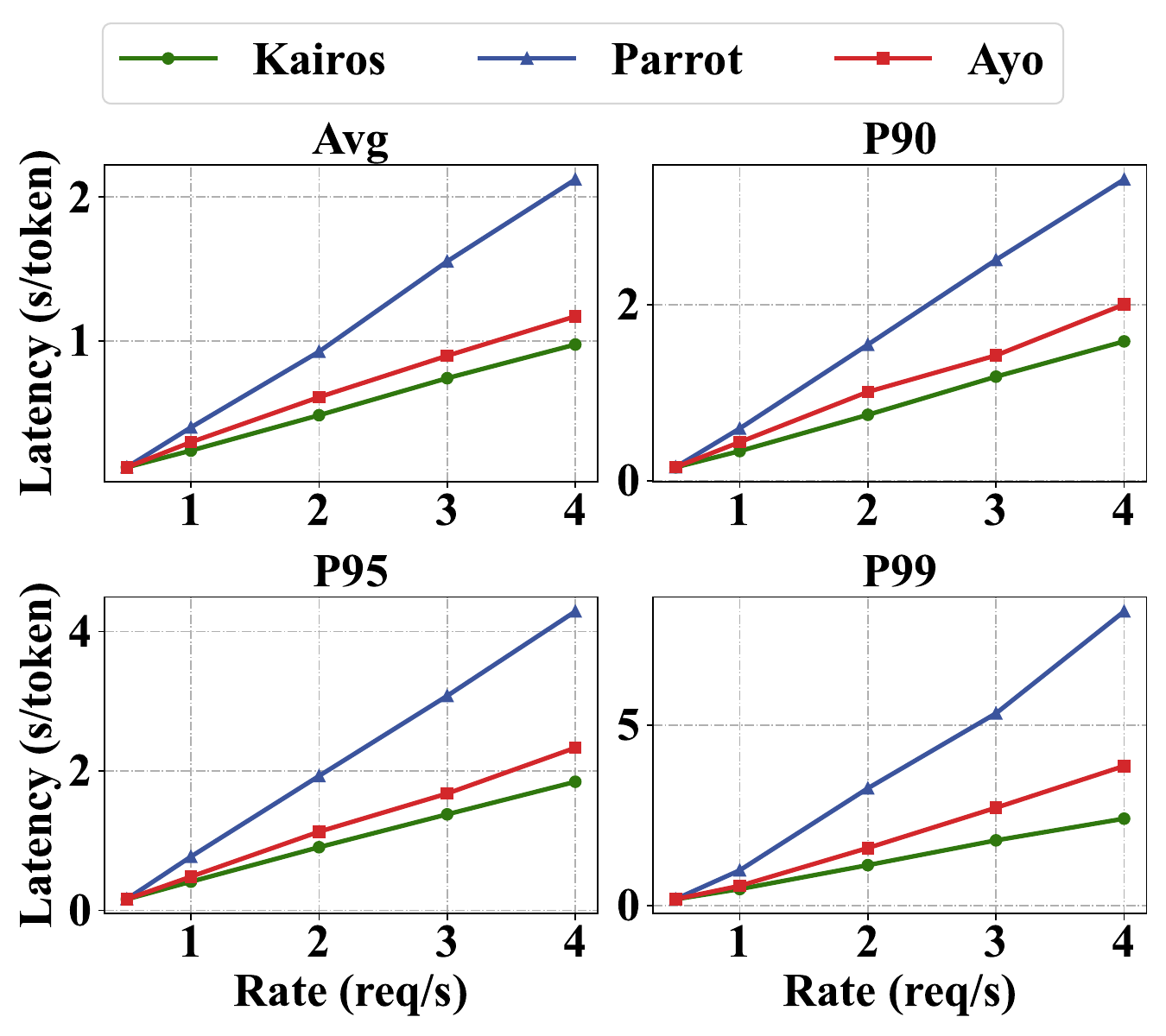}
    \caption{Latency performance of co-located applications (QA, RG, and CG) using Llama2-13B.}
  \label{fig:colocation_latency_13B}
  \vspace{-3mm}
\end{figure}

As shown in Figure~\ref{fig:colocation_latency_13B}, \system{} achieves significant performance improvements over both Parrot and Ayo. Compared to Parrot, \system{} reduces the average latency by 42.1\% to 57.4\%, P90 latency by 43.1\% to 56.6\%, P95 latency by 48.1\% to 57.1\%, and P99 latency by 56.8\% to 70.6\%. Moreover, compared to Ayo, \system{} reduces the average latency by 21.8\% to 24.6\%, P90 latency by 23.2\% to 25.9\%, P95 latency by 18.4\% to 21.4\%, and P99 latency by 22.6\% to 37.9\%. These results demonstrate that \system{} continues to maintain a significant performance advantage under a larger LLM, showcasing excellent scalability and adaptability.

\subsection{Ablation Study}

This section reuses the same workload and system configuration as in Section~\ref{sec:co-location}, aiming to validate the effectiveness of \system{}'s two core mechanisms. We construct the following two ablation variants:

\begin{itemize}
    \item \textbf{w/o priority}: Removes the priority scheduling strategy, retaining the request packing strategy.
    \item \textbf{w/o packing}: Removes the request packing strategy, retaining the priority scheduling strategy.
\end{itemize}

\begin{figure}
  \centering
  \includegraphics[width=0.85\linewidth]{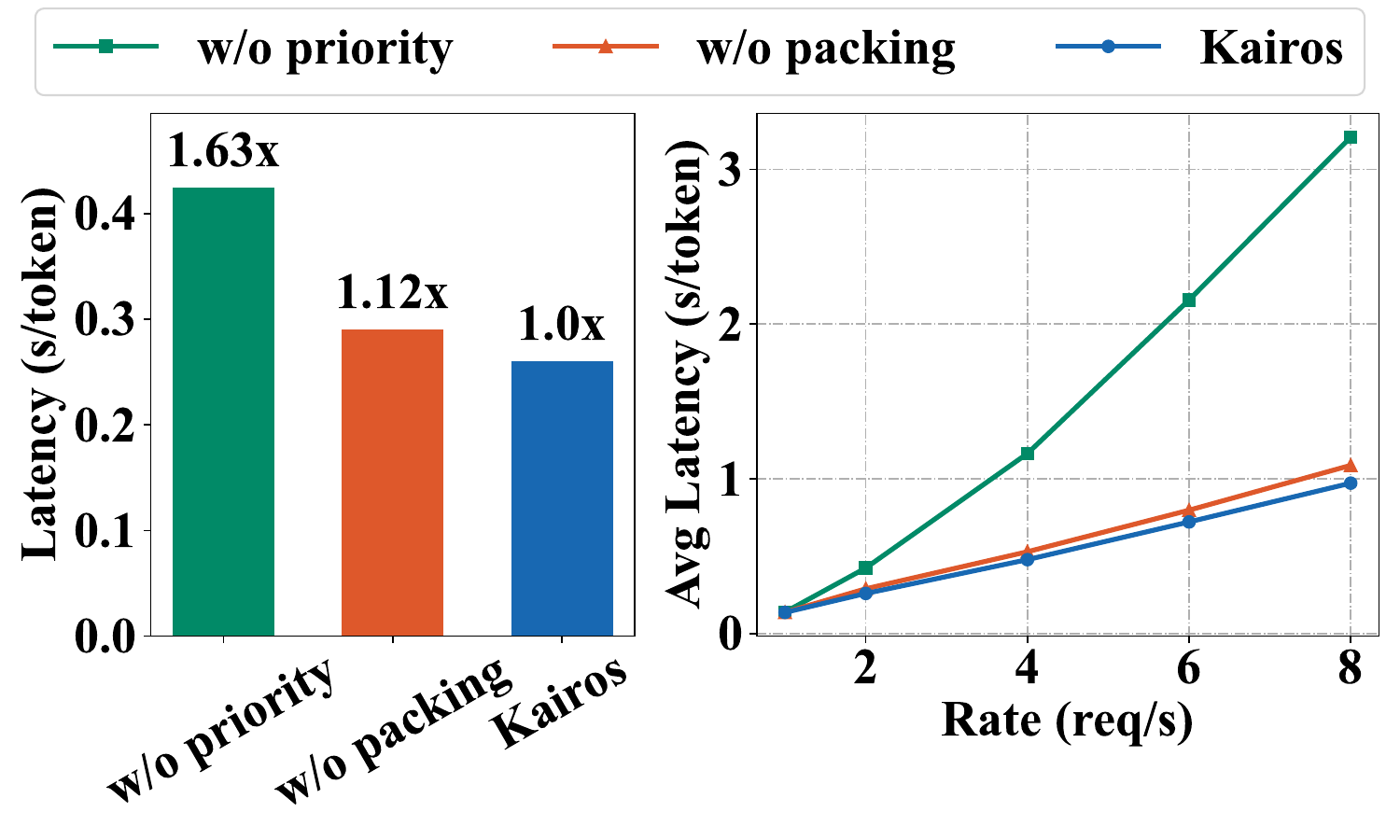}
    \caption{Results of the ablation studies.}
  \label{fig:abl_all}
  \vspace{-3mm}
\end{figure}

\textbf{Effectiveness of workflow-aware priority scheduling.} As shown in Figure~\ref{fig:abl_all}, we compare the performance of \system{} to its variant \textit{w/o priority} under different request rates.
The left subfigure illustrates the average latency under a representative load where queueing time accounts for 50\%. The results indicate that \system{} can mitigate the impact of queueing blocking on system performance through priority scheduling, and achieves a 1.63x performance improvement.
The right subfigure presents the average latency at different request rates. As the request rate increases, \system{}'s latency improvement increases from 38.8\% to 69.6\%. This is because request-level head-of-line blocking becomes more frequent under higher loads, while \system{}'s workflow-aware priority scheduling strategy can effectively alleviate such blocking, thereby significantly improving overall system performance.

\textbf{Effectiveness of memory-aware time-slot dispatching.} As shown in Figure~\ref{fig:abl_all}, we compare the performance of \system{} with its variant \textit{w/o packing} under different request rates. 
The left subfigure indicates that \system{} achieves a 1.12x acceleration because its request packing strategy can dynamically match appropriate LLM instances based on the memory demands of different agent requests, effectively reducing resource waste and request preemption. 
The right subfigure presents the average latency at different request rates. \system{} achieves a stable latency reduction ranging from 9.5\% to 10.6\%. The main reason is that most of the GPU memory is effectively utilized by requests, and the remaining wasted memory space changes little across different request rates. This further demonstrates that \system{}'s request dispatching exhibits stable and reliable optimization capabilities.

\subsection{Overhead of \system{}}
We analyze the overhead from two dimensions: real-time updates of agent-level priority and request execution.

Firstly, \system{} computes and updates agent priorities using real-time latency distributions based on Wasserstein distance and MDS methods. When calculating the Wasserstein distance matrix, \system{} only requires incremental computation for newly added agents, which results in negligible overhead. The MDS method exhibits quadratic computational complexity with respect to the number of agents~\cite{6683892,7850955}. We evaluate its computation time for agent scales from 10 to 5000, ranging approximately from 0.1s to 4.3s, which is overall within an acceptable range. These operations can be triggered at fixed time intervals and executed asynchronously in the background, ensuring high scalability and performance in large-scale agent scenarios.

Additionally, during the execution of each request, \system{} introduces two types of additional overhead. First, the priority scheduling strategy needs to sort queued requests, which takes approximately 3.6 ms. Second, the request packing strategy needs to compute GPU memory usage based on time slots, taking approximately 4.1 ms. These overheads are negligible compared to the latency of LLM inference. 

\section{Related Work}
\textbf{LLM Serving.} Extensive research has been dedicated to optimizing general LLM inference performance across various directions, including techniques such as continuous batching~\cite{280922,kwon2023efficient}, KV cache management~\cite{lin2024infinitellmefficientllmservice, sheng2024sloraservingthousandsconcurrent, kwon2023efficient,zheng2024sglangefficientexecutionstructured}, request scheduling~\cite{298679,10.5555/3691938.3691990}, kernel accelerating~\cite{10.5555/3600270.3601459,10.5555/3618408.3619993}, and model parallelism~\cite{288582,10.1145/3620665.3640411}. For multi-instance load balancing, techniques like live migration~\cite{298685} and disaggregated prefill and decoding~\cite{298687,305212} have been proposed to improve end-to-end performance.
While these methods significantly enhance the inference performance of LLMs themselves, they are limited in optimizing multi-agent applications as they cannot leverage application-level information for end-to-end performance optimization. \system{}, however, takes a holistic view by focusing on end-to-end optimization for multi-agent applications. These LLM inference layer optimizations are orthogonal to \system{} and can be seamlessly integrated.

\textbf{LLM orchestration frameworks.} To address the complex inter-agent interactions and tool execution within multi-agent applications, frameworks like LangChain~\cite{langchain} and AutoGen~\cite{wu2023autogenenablingnextgenllm} have emerged. These frameworks primarily offer high-level programming abstractions for client-side workflow definition to simplify user development, but they do not incorporate LLM serving optimizations essential for end-to-end application performance~\cite{crewAI,langgraph,camle,langflow,zhou2023agentsopensourceframeworkautonomous}. Recognizing this gap, \system{} also provides a front-end orchestrator to simplify user programming while concurrently enabling the automated collection of necessary information for LLM serving performance optimization.

\textbf{Optimization for applications.} Several works have been proposed to optimize LLM-based applications. 
Parrot~\cite{10.5555/3691938.3691988} is an LLM service system optimizing end-to-end performance of LLM-based applications with semantic variable, adopting a FCFS scheduling strategy and a Round-Robin dispatching strategy.
Ayo~\cite{10.1145/3676641.3716278} is a fine-grained orchestration framework, which introduces a priority scheduling strategy based on workflow topology depth, and dispatches requests in a Round-Robin manner.
However, they fail to account for the diverse execution characteristics of agents in multi-agent applications and lack an effective orchestrator to automatically collect and parse such information. This deficiency leads to inefficient scheduling and dispatching of multi-agent application requests under excessive load.

\section{Conclusion}
In this paper, we present \system{}, a scalable multi-agent orchestration system for multi-agent applications. 
The core idea is to optimize scheduling and dispatching by leveraging the execution differences among agent requests through three main components: a workflow orchestrator that automatically collects agent-specific information, a workflow-aware scheduler that optimizes priority decisions by perceiving the remaining latency of requests across agents, and a memory-aware dispatcher that efficiently dispatches requests to LLM instances by considering varying memory demands among agents.
Experimental results show that \system{} reduces the end-to-end latency by 17.8\% to 28.4\%, compared to state-of-the-art works.

\bibliographystyle{ACM-Reference-Format}
\bibliography{sample-sigplan}

\end{document}